\documentclass[aps,twocolumn,superscriptaddress,floatfix,nofootinbib]{revtex4-1}
\usepackage{graphicx,amsmath,amssymb,verbatim,color}
\usepackage{booktabs}
\usepackage{comment}
\usepackage{soul}
\usepackage[dvipsnames]{xcolor}
\usepackage{bm}
\usepackage[utf8]{inputenc}
\usepackage[colorlinks=true,citecolor=blue,linkcolor=blue,urlcolor=blue, backref=false,pdfborder={0 0 0}]{hyperref}
\usepackage{float}
\usepackage{multirow}
\newcommand{\be}{\begin{equation}\begin{gathered}}
\newcommand{\ee}{\end{gathered}\end{equation}} 
\newcommand{\barr}{\begin{eqnarray}}
\newcommand{\earr}{\end{eqnarray}} 
\usepackage{romannum}
\usepackage[normalem]{ulem}

\usepackage{makecell}
\usepackage{romannum}

\begin{document}

\pagenumbering{arabic}

\title{What it takes to solve the Hubble tension\\through Modifications of Cosmological Recombination \Romannum{2}:\\ in light of ACT DR6 and DESI DR2}

\author{Nanoom Lee}
\email{nanoom.lee@jhu.edu}
\affiliation{William H. Miller III Department of Physics \& Astronomy, Johns Hopkins University, Baltimore, Maryland 21218, USA}
\author{Tianji Zhou}
\affiliation{Department of Physics and Astronomy, Haverford College, Haverford, Pennsylvania 19041, USA}

\date{\today} 
\begin{abstract}
We construct data-driven solutions to the Hubble tension, in light of recent data from the Atacama Cosmology Telescope (ACT DR6) and the Dark Energy Spectroscopic Instrument (DESI DR2). We search for the minimal modification to the recombination history through a time-varying electron mass $m_e(z)$ that increases the best-fit $H_0$ inferred from CMB data toward the SH0ES value, without worsening the fit to the data. Using \textit{Planck} and ACT data including lensing, we find a perturbative modification to $m_e(z)$ that fully resolves the Hubble tension, with the solution sharing the same qualitative oscillatory structure as in previous work using \textit{Planck} data alone, demonstrating its robustness to the inclusion of more precise and independent CMB data. As a byproduct, the solution also eases the $S_8$ tension. Once DESI DR2 BAO data are added, however, perturbative modifications to $m_e(z)$ cannot fully resolve the Hubble tension. This reflects the same fundamental limitation: raising $H_0$ by modifying recombination generically lowers $\Omega_m$, being inconsistent with late-time cosmological observations.
\end{abstract}

\maketitle

\section{Introduction}
\label{sec:intro}

The Hubble tension --- the persistent discrepancy between the value of the Hubble constant inferred from early-Universe observations and that measured using late-time distance indicators --- remains one of the most significant open problems in modern cosmology. The most well-known early-Universe measurement is that inferred from the \textit{Planck} satellite's cosmic microwave background (CMB) anisotropy data, $H_0 = 67.36 \pm 0.54\;\mathrm{km\,s^{-1}\,Mpc^{-1}}$ \cite{Planck:2018vyg}, while the local measurement is provided by the SH0ES collaboration, $H_0 = 73.04 \pm 1.04\;\mathrm{km\,s^{-1}\,Mpc^{-1}}$ \cite{Riess:2021jrx}, recently updated to $H_0 = 73.17 \pm 0.86\;\mathrm{km\,s^{-1}\,Mpc^{-1}}$ with an improved distance ladder calibration \cite{Breuval:2024lsv}. The discrepancy between these two values now exceeds $5\sigma$ \cite{Verde:2019ivm}, making it unlikely to be a statistical fluctuation and motivating extensive scrutiny of both systematic uncertainties and extensions to the standard cosmological model \cite{Murgia:2016ccp, Pourtsidou:2016ico, Nunes:2016dlj, DiValentino:2016hlg, Kumar:2016zpg, Karwal:2016vyq, Kumar:2017dnp, DiValentino:2017iww, DiValentino:2017rcr, DiValentino:2017oaw, Dutta:2018vmq, Yang:2018uae, Poulin:2018cxd, DiValentino:2019exe, Visinelli:2019qqu, Pan:2019hac, DiValentino:2019ffd, Smith:2019ihp, Knox:2019rjx, Arendse:2019hev, Niedermann:2019olb, vonMarttens:2019ixw, Akarsu:2019hmw, Yang:2020uga, Perez:2020cwa, Ye:2020btb, Yang:2020zuk, Lucca:2020zjb, Krishnan:2020obg, Blinov:2020hmc, Jedamzik:2020krr, DiValentino:2020naf, Calderon:2020hoc, DiValentino:2020leo, DiValentino:2020vnx, Yang:2021flj, Kumar:2021eev, Yang:2021eud, Nunes:2021zzi, Ye:2021iwa, Akarsu:2021fol, Poulin:2021bjr, DiValentino:2021rjj, Alestas:2021luu, Gariazzo:2021qtg, Niedermann:2021vgd, Sen:2021wld, Heisenberg:2022lob, Anchordoqui:2022gmw, DiGennaro:2022ykp, Schoneberg:2022grr, Yao:2022kub, Akarsu:2022typ, Ong:2022wrs, Lee:2022gzh, Bernui:2023byc, Mishra:2023ueo, vanderWesthuizen:2023hcl, deSouza:2023sqp, Poulin:2023lkg, Zhai:2023yny, Giare:2023xoc, Cruz:2023lmn, Adil:2023exv, Ruchika:2023ugh, Greene:2023cro, Liu:2023kce, Niedermann:2023ssr, Frion:2023xwq, Akarsu:2023mfb, Hoerning:2023hks, Vagnozzi:2023nrq, Gomez-Valent:2023uof, Pan:2023mie, Lapi:2023plb, Castello:2023zjr, Efstathiou:2023fbn, Cervantes-Cota:2023wet, Forconi:2023hsj, Anchordoqui:2023woo, Pan:2023frx, Garcia-Arroyo:2024tqq, Akarsu:2024qsi, Paraskevas:2024ytz, Benisty:2024lmj, Silva:2024ift, Manoharan:2024thb, Krolewski:2024jwj, Halder:2024uao, Greene:2024qis, Giare:2024ytc, Garny:2024ums, Giare:2024akf, Anchordoqui:2024gfa, Gomez-Valent:2024tdb, Allali:2024anb, Bagherian:2024obh, Bousis:2024rnb, Baryakhtar:2024rky, Seto:2024cgo, Co:2024oek, Akarsu:2024eoo, Yadav:2024duq, Lynch:2024hzh, Aboubrahim:2024spa, Tang:2024gtq, Poulin:2024ken, Toda:2024ncp, Schoneberg:2024ynd, Jiang:2024xnu, Pedrotti:2024kpn, Kochappan:2024jyf, Ruchika:2024ymt, Simon:2024jmu, Mirpoorian:2024fka, Gomez-Valent:2024ejh, Souza:2024qwd, Sabogal:2025mkp, Akarsu:2025gwi, Soriano:2025gxd, Akarsu:2025ijk, Silva:2025hxw, Escamilla:2025imi, Smith:2025zsg, Scherer:2025esj, Specogna:2025guo, Mirpoorian:2025rfp, Bouhmadi-Lopez:2025ggl, Lee:2025pzo, Lee:2025yah, Toda:2025kcq, GarciaEscudero:2025lef, Efstratiou:2025iqi, Ghafari:2025eql, Smith:2025uaq, Kumar:2025obb, Zhou:2025kws}.

Among early-Universe solutions, modifications to the cosmological recombination history constitute a particularly well-motivated possibility. Recombination directly determines both the sound horizon $r_d$ and the detailed structure of the CMB anisotropy spectra. Reducing $r_d$ through modifications to recombination naturally leads to a larger inferred $H_0$, in order to keep the angular scale of the sound horizon $\theta_s$ fixed at its precisely measured value. Ref.~\cite{Lee:2022gzh} (hereafter Paper~\Romannum{1}) introduced a data-driven, model-independent framework to search for extensions to $\Lambda$CDM that can reconcile CMB-inferred parameters with external measurements, without degrading the fit to the data, and applied it to modifications of the recombination history. Without any model-building effort, this approach uses the Fisher-bias formalism to identify perturbations to the recombination history that achieve a targeted shift in parameters such as $H_0$.\footnote{This method is generic and has been applied in different cosmological contexts \cite{Mirpoorian:2024fka, Zhou:2025kws, Lee:2025yah}.} When applied to \textit{Planck} CMB data, this framework demonstrated that modest, perturbative changes to a time-varying electron mass $m_e(z)$ could, in principle, raise the inferred value of $H_0$ to match local measurements without worsening the fit to CMB anisotropies alone. However, the inclusion of additional late-time datasets --- baryon acoustic oscillations (BAO) and uncalibrated supernovae --- significantly restricted the viability of such solutions, due to the generic tendency of recombination-based solutions to lower $\Omega_m$. These results highlighted both the potential and the limitations of such solutions, while providing a quantitative benchmark for what any concrete physical model would need to accomplish.

In recent years, high-precision ground-based CMB experiments have delivered independent measurements of the CMB anisotropy spectra, probing smaller angular scales than \textit{Planck} with independent instrumental systematics. In particular, the Atacama Cosmology Telescope (ACT) has recently released its sixth data release (DR6) \cite{AtacamaCosmologyTelescope:2025blo}, providing CMB temperature and polarization power spectra measured to arcminute scales over $\sim\!19{,}000\;\mathrm{deg}^2$ of sky. From CMB data alone, the joint \textit{Planck}+ACT dataset yields $H_0 = 67.62 \pm 0.5\;\mathrm{km\,s^{-1}\,Mpc^{-1}}$~\cite{AtacamaCosmologyTelescope:2025blo}, consistent with \textit{Planck} and in tension with SH0ES at the $\sim\!5\sigma$ level. ACT's sensitivity to high-multipole temperature and polarization anisotropies makes it especially relevant for testing recombination physics, which leaves characteristic imprints on the acoustic peak structure and damping tail of the CMB power spectra. Complementary to this, the Dark Energy Spectroscopic Instrument (DESI) has recently released its second data release (DR2) \cite{DESI:2025zgx}, providing BAO measurements of significantly improved precision over previous surveys such as BOSS DR12 \cite{BOSS:2016wmc}, across a wide redshift range with multiple tracer populations.

In this work, we apply the data-driven framework of Paper~\Romannum{1} to these new datasets --- ACT DR6 CMB and DESI DR2 BAO --- to assess recombination-based solutions to the Hubble tension. The remainder of this paper is organized as follows. In Sec.~\ref{sec:method}, we briefly review the framework searching for data-driven solutions to the Hubble tension. Section~\ref{sec:data} describes the datasets considered in this work. Our results are presented in Sec.~\ref{sec:results}, followed by a discussion in Sec.~\ref{sec:discussion}.

\section{Methodology}
\label{sec:method}

This work follows the method developed in Paper~\Romannum{1} \cite{Lee:2022gzh}, to which we refer the reader for a detailed derivation, and provide only a brief summary here.

We start with a fiducial flat $\Lambda$CDM cosmology described by six parameters $\vec{\Omega} = \left\{\omega_b,\omega_c,H_0,\tau,A_s,n_s\right\}$, where $\omega_b$ and $\omega_c$ are the baryon and cold dark matter densities, $H_0$ is the Hubble constant, $\tau$ is the optical depth to reionization, and $A_s$ and $n_s$ are the amplitude and tilt of the primordial scalar power spectrum. We denote the observed data by $\mathbf{X}^{\rm obs}$ and the theoretical prediction by $\mathbf{X}(\vec{\Omega})$, where $\mathbf{X}$ may contain any cosmological observables. Both the best-fit parameters $\vec{\Omega}_{\rm BF}$ and the best-fit chi-squared $\chi^2_{\rm BF}$ depend on the underlying theoretical model, and will change when the model is perturbed.

We consider perturbations $\Delta f(z)$ to a smooth function $f(z)$ on which the model depends. Our goal is to find such perturbations that shift the best-fit $H_0$ toward the SH0ES value without worsening the fit to the data. We stress that our primary goal is not to find a compelling physical solution, but to establish whether such solutions exist. This is cast as the constrained optimization problem:
\begin{equation}
\mathrm{minimize}\left(\|\Delta f\|^2\right)
\quad \textrm{with} \quad
\left\{
\begin{aligned}
&H_{0}^{\mathrm{BF}}[\Delta f(z)] = H_{0}^{\mathrm{target}},\\
&\Delta \chi^2_{\mathrm{BF}}[\Delta f(z)] \le 0,
\end{aligned}
\right.
\label{eq:minimization}
\end{equation}
where $\|\Delta f\|^2 \equiv \int dz\,[\Delta f(z)]^2$. While this problem could in principle be solved with MCMC, that would be computationally prohibitive. Instead, we use the Fisher-bias formalism \cite{Knox:1998fp,Kim:2003mq,Taylor:2006aw,Shapiro:2008yk,DeBernardis:2008tk} to obtain analytic approximations for $\Delta H_0^{\rm BF}$ and $\Delta\chi^2_{\rm BF}$ when $\Delta f(z)$ is small. Specifically, by Taylor-expanding $\chi^2$ to second order around the fiducial cosmology, one obtains
\barr
\Delta \Omega_{\rm BF}^i &=& \int dz\; \frac{\delta \Omega_{\rm BF}^i}{\delta  f(z)} \Delta f(z),\label{eq:DObf_X}\\
\Delta \chi_{\rm BF}^2 &=& \int dz\; \frac{\delta \chi_{\rm BF}^2}{\delta f(z)}\Delta f(z) \nonumber\\
&+& \frac12 \iint dz\;dz'\; \frac{\delta^2 \chi_{\rm BF}^2}{\delta f(z) \delta f(z')} \Delta f(z) \Delta f(z'),~~\label{eq:Dchi2bf_X}
\earr
where the functional derivatives are given by
\barr
\frac{\delta \Omega_{\rm BF}^i}{\delta f(z)} &=& -\left(F^{-1}\right)_{ij} \frac{\partial \bm{X}}{\partial \Omega^j} \cdot \bm{M}\cdot\frac{\delta \bm{X}}{\delta f(z)},\label{eq:dObf}\\
\frac{\delta \chi_{\rm BF}^2}{\delta f(z)} &=& 2\left[\bm{X}(\vec{\Omega}_{\rm fid}) - \bm{X}^{\rm obs}\right] \cdot\widetilde{\bm{M}} \cdot \frac{\delta \bm{X}}{\delta f(z)},\label{eq:dchi2bf_lin}\\
\frac{\delta^2 \chi_{\rm BF}^2}{\delta f(z) \delta f(z')} &=& 2 \frac{\delta \bm{X}}{\delta f(z)}\cdot\widetilde{\bm{M}}\cdot\frac{\delta \bm{X}}{\delta f(z')},\label{eq:dchi2bf_quad}
\earr
with $F_{ij}$ the Fisher matrix, $\mathbf{M}$ the inverse covariance matrix, 
and
\be
\widetilde{M}_{\alpha\beta} \equiv M_{\alpha\beta} - M_{\alpha \gamma} \frac{\partial X^\gamma}{\partial \Omega^i}(F^{-1})_{ij}\frac{\partial X^\sigma}{\partial \Omega^j}M_{\sigma\beta}.
\ee
With these expressions, the optimization problem of Eq.~\eqref{eq:minimization} becomes tractable. We refer to Paper~\Romannum{1} \cite{Lee:2022gzh} for the derivation of these expressions. Note that Eq.~\eqref{eq:dObf} can be interpreted as computing the part of the signal $\delta\bm{X}/\delta f(z)$ that is degenerate with shifts in the standard cosmological parameters $\vec{\Omega}$, through the projection onto the subspace spanned by $\partial\bm{X}/\partial\Omega^i$. This is analogous to the generalized Fisher forecast of Ref.~\cite{Lee:2021bmn}, where the same projection was used to isolate the non-degenerate part of a physical signal with respect to standard cosmological parameters.

We apply this framework to a time-varying electron mass, taking $f(z) = \ln m_e(z)$ following Paper~\Romannum{1}. For large target shifts in $H_0$, we apply the method iteratively following Ref.~\cite{Lee:2025yah}. The validity of the Fisher-bias approach was demonstrated in Paper~\Romannum{1}, where the estimated parameter shifts were shown to agree with full MCMC results at the $\sim\!1\sigma$ level for all six cosmological parameters.

We note that varying $m_e(z)$ modifies the recombination history through its effect on the atomic physics of hydrogen and helium \cite{Kaplinghat:1998ry,Scoccola:2009xtv,Planck:2014ylh,Chluba:2015gta,Hart:2017ndk}. Specifically, it rescales the Einstein coefficients, photoionization and recombination rates, the
two-photon decay rate, the Thomson scattering cross section ($\sigma_T \propto m_e^{-2}$), and the effective temperature, following the dependencies summarized in Appendix~B of Paper~\Romannum{1} and implemented in \texttt{HYREC-2} \cite{Ali-Haimoud:2010tlj, Ali-Haimoud:2010hou, Lee:2020obi}. As a result, a time-varying $m_e(z)$ is not equivalent to a direct modification of the free-electron fraction $x_e(z)$ alone, since $\sigma_T$ also changes, affecting CMB diffusion damping independently. While $\sigma_T$ and $x_e$ are partially degenerate through the photon scattering rate $\Gamma_T \propto x_e \sigma_T$, their redshift dependences differ when $m_e$ varies.

The functional derivatives are evaluated numerically using $N = 150$ equally-spaced narrow Gaussian perturbations to $f(z)$ in the range $z \in [300, 3000]$, following Ref.~\cite{Lee:2022gzh}\footnote{In practice, the covariance matrix of the data is not always directly accessible for all likelihood functions. We therefore compute all required quantities via numerical derivatives of the likelihood functions, without explicitly constructing the covariance matrix. This procedure is described in detail in Appendix~\ref{appendix:method}.}:
\be
\frac{\Delta m_e}{m_e}(z, z_i) \propto 
\exp\left[-\frac{(z - z_i)^2}{2\sigma^2}\right], \quad
\sigma \equiv \frac{z_{\max} - z_{\min}}{6N\sqrt{2\ln 2}}.
\ee
The CMB spectra are computed using a modified version of \texttt{CLASS v3.3.1} \cite{class}, with \texttt{HYREC-2} \cite{Ali-Haimoud:2010tlj, Ali-Haimoud:2010hou, Lee:2020obi} implemented for the recombination history.

\section{Data}
\label{sec:data}

\textbf{Low-$\ell$ \textit{Planck}.}
For low-$\ell$ ($\ell < 30$) TT and EE spectra, we adopt the compressed Gaussian likelihood code from Ref.~\cite{Lee:2026toh}\footnote{Available at \href{https://github.com/nanoomlee/planck-gaussian-lowl}{github.com/nanoomlee/planck-gaussian-lowl}.}, which contains the compressed low-$\ell$ EE likelihood constructed from the \texttt{Sroll2} likelihood for \textit{Planck} low-$\ell$ EE data \cite{Planck:2016kqe,Pagano:2019tci} and the compressed low-$\ell$ TT likelihood from Ref.~\cite{Prince:2021fdv} based on \textit{Planck} \texttt{Commander} likelihood \cite{Eriksen:2007mx}. The \texttt{Sroll2} likelihood \cite{Planck:2016kqe,Pagano:2019tci} provides the tightest constraint on $\tau$ to date and has been used as the default low-$\ell$ EE likelihood in the ACT DR6 analysis. In both likelihoods, the likelihood for binned spectra $D_\ell \equiv \ell(\ell+1)C_\ell/2\pi$ takes a log-normal form,
\be
\mathcal{L}(D) = p(D) = \frac{1}{(D-D_0)\sigma\sqrt{2\pi}} e^{-[\ln(D-D_0)-\mu]^2/(2\sigma^2)},
\ee
where $D = D_{\rm bin}$ with two bins for the TT \cite{Prince:2021fdv} and six bins for the EE spectra \cite{Lee:2026toh}, respectively, and the values of $D_0$, $\mu$, and $\sigma$ are determined in Refs.~\cite{Prince:2021fdv,Lee:2026toh}. The chi-squared from this likelihood is
\barr
\chi^2_{\text{low-}\ell\,{\rm TT\;or\;EE}} &\equiv& -2\sum_i^{N_{\rm bin}}  \ln\mathcal{L}(D_i) \nonumber \\
&=& \sum_i^{N_{\rm bin}}\frac{[\ln(D_i-D_{0,i}) - \mu_i + \sigma_i^2]^2}{\sigma_i^2},~~~
\earr
up to a constant. Since this has the same form as a Gaussian $\chi^2$, our data vector $\mathbf{X}$ includes $\ln(D_i - D_{0,i})$ as its components rather than $D_i$ itself. This treatment has been validated in Ref.~\cite{Lee:2026toh}.\\

\textbf{\textit{Planck}-lite and ACT-lite (high-$\ell$).}
For high-$\ell$ CMB temperature and polarization data, we use the \textit{Planck}-lite \cite{Planck2018} and ACT-lite \cite{AtacamaCosmologyTelescope:2025blo} likelihoods. These are foreground-marginalized likelihoods, with scale cuts and treatment of the overlapping sky region between \textit{Planck} and ACT as defined in Ref.~\cite{AtacamaCosmologyTelescope:2025blo}.\\

\textbf{CMB lensing.}
We include CMB lensing measurements from the ACT DR6 lensing likelihood \cite{ACT:2023kun,ACT:2023dou,Carron:2022eyg}, which provides a combined likelihood for \textit{Planck} + ACT DR6 lensing power spectrum \texttt{Planck-ACT-baseline}.\footnote{\href{https://github.com/ACTCollaboration/act_dr6_lenslike}{https://github.com/ACTCollaboration/act\_dr6\_lenslike}} While we include the lensing data for the consistency with the ACT baseline analysis, as demonstrated in Paper~\Romannum{1} \cite{Lee:2022gzh}, the lensing data have no significant impact on the recombination solutions; we have verified that this remains the case with the ACT DR6 lensing data.\\

\textbf{BAO -- DESI DR2.}
We use DESI DR2 BAO measurements \cite{DESI:2025zgx}, which provide anisotropic measurements of the comoving angular diameter distance $D_M(z_{\rm eff})$ and the Hubble rate $H(z_{\rm eff})$, both scaled by the sound horizon at the baryon drag epoch $r_d$, across multiple tracer populations spanning $0.1 \lesssim z_{\rm eff} \lesssim 2.3$.\footnote{Available at \href{https://github.com/CobayaSampler/bao_data/tree/master/desi_bao_dr2}{https://github.com/CobayaSampler/bao\_data/tree/\allowbreak master/desi\_bao\_dr2}.} When computing the functional derivatives $\delta\mathbf{X}/\delta\ln m_e(z)$, we neglect the response of the BAO observables to perturbations in $m_e(z)$, since this effect is subdominant compared to that on the CMB anisotropy spectra for the weak perturbations considered here, as we have verified numerically.\\

\textbf{Datasets.}
We consider two datasets in this work. The first, \textbf{P-ACT-L}, combines the low-$\ell$ \textit{Planck} likelihood, the \textit{Planck}-lite and ACT-lite high-$\ell$ likelihoods, and the \textit{Planck} + ACT DR6 CMB lensing. The label ``L'' denotes the inclusion of CMB lensing. The second, \textbf{P-ACT-LB}, additionally includes DESI DR2 BAO (``B'').

\section{Results}
\label{sec:results}

We present the results of applying the Fisher-bias framework to two datasets: P-ACT-L (\textit{Planck}+ACT CMB including lensing) and P-ACT-LB (P-ACT-L combined with DESI DR2 BAO). In both cases, we search for the minimal time-varying electron mass $m_e(z)$ that shifts the best-fit $H_0$ to a target value while not worsening the fit to the data, solving Eq.~\eqref{eq:minimization} with $f(z)\equiv \ln m_e(z)$. Unlike Paper~I, we do not include uncalibrated supernovae data, as we have verified that their inclusion does not qualitatively change our conclusions given the tight constraints on $\Omega_m$ already provided by DESI DR2 BAO.

\subsection{Application to P-ACT-L data}

\begin{figure}[t!]
\centering
\includegraphics[width = \linewidth,trim= 00 20 00 0]{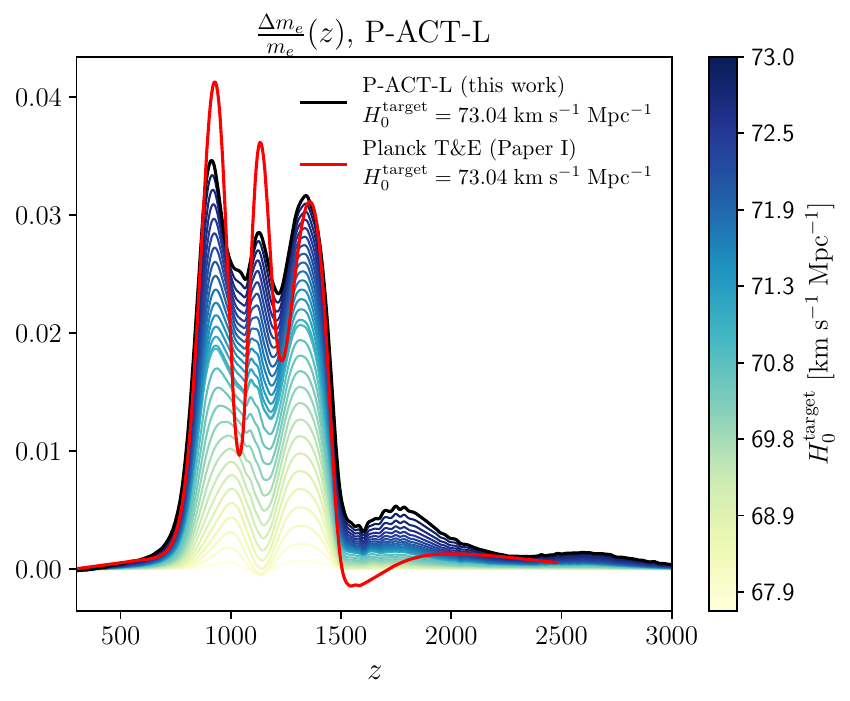}
\caption{Solutions for $\Delta m_e/m_e(z)$ constructed to achieve a target best-fit Hubble constant $H_0^{\rm target}$ using the P-ACT-L dataset, shown for a range of target values from $H_0^{\rm target} \approx 67.67$ to $73.04\;\mathrm{km\,s^{-1}\,Mpc^{-1}}$ (color scale). All solutions are constructed to leave the best-fit $\chi^2$ unaffected. The black curve corresponds to $H_0^{\rm target} = 73.04\;\mathrm{km\,s^{-1}\,Mpc^{-1}}$, consistent with the SH0ES measurement \cite{Riess:2021jrx}. For comparison, the red curve shows the corresponding solution from Paper~\Romannum{1} obtained with \textit{Planck} T\&E data\footnote{This includes \textit{Planck}-lite with $\ell_{\rm max}=2508$ for temperature and $\ell_{\rm max}=1996$ for polarizations, and the compressed low-$\ell$ \textit{Planck} likelihood of Ref.~\cite{Prince:2021fdv} for both TT and EE.} alone.}
\label{fig:P-ACT-L-solusions}
\end{figure}

Figure \ref{fig:P-ACT-L-solusions} shows the solutions $\Delta m_e/m_e(z)$ constructed for a range of target values $H_0^{\rm target}$, using the P-ACT-L dataset. For large target shifts in $H_0$, we apply the method iteratively following Ref.~\cite{Lee:2025yah}. Specifically, for target values $H_0^{\rm target} > 71\;\mathrm{km\,s^{-1}\,Mpc^{-1}}$, we first construct a solution with $H_0^{\rm target} = 71\;\mathrm{km\,s^{-1}\,Mpc^{-1}}$, update the fiducial cosmology to the new best-fit, and then construct a second solution on top of the first to achieve the desired target value. We have verified that two iterations are sufficient for convergence. The solution targeting $H_0^{\rm target} = 73.04\;\mathrm{km\,s^{-1}\,Mpc^{-1}}$ (the SH0ES best-fit value~\cite{Riess:2021jrx}) is highlighted (black). As in Paper~\Romannum{1}, the solution exhibits oscillatory behavior concentrated in the redshift range $z \sim 700$--$1500$, where modifications to the recombination history have the strongest impact on CMB anisotropy power spectra. Notably, despite ACT DR6 providing substantially more precise measurements of the CMB damping tail at small angular scales with independent instrumental systematics compared to \textit{Planck}, a solution fully resolving the Hubble tension can still be found. The shape of this solution is in close agreement with the \textit{Planck} T\&E solution of Paper~\Romannum{1} (red), demonstrating that the recombination-based solution is robust to the inclusion of a more precise and independent CMB dataset. We note, however, that the intermediate trough between the two main peaks around $z \sim 1100$--$1200$ is notably shallower in the P-ACT-L solution compared to Paper~\Romannum{1}; we discuss this further in Appendix~\ref{appendix:dchi2}.

\begin{figure*}[ht!]
\centering
\includegraphics[width = \linewidth,trim= 00 10 00 10]{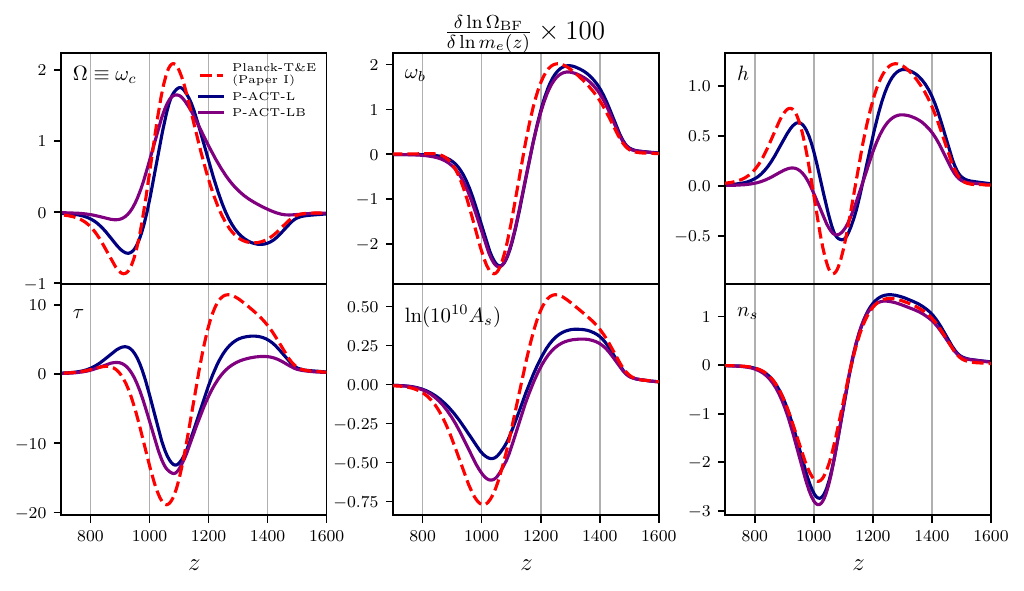}
\caption{Functional derivatives of best-fit parameters with respect to fractional changes in the electron mass, $\delta\ln\Omega_{\rm BF}^i/\delta\ln m_e(z)$ [Eq.~\eqref{eq:dObf}], for three datasets: \textit{Planck} T\&E (red dashed, from Paper~\Romannum{1} \cite{Lee:2022gzh}), P-ACT-L (navy), and P-ACT-LB (purple). All datasets share the same qualitative oscillatory structure. 
Apparent differences in amplitude for individual parameters such as $\tau$, $\ln(10^{10}A_s)$, $\omega_c$, and $h$ reflect the different ways in which each dataset breaks the standard CMB parameter degeneracies --- most notably the $\tau$--$A_s$ degeneracy and the $\Omega_m$--$h$ degeneracy through $\Omega_m h^3$ --- rather than any physically distinct response to $m_e(z)$ (see Fig.~\ref{fig:degeneracy} below, and Fig.~6 of Ref.~\cite{Lee:2022gzh}). See Appendix~\ref{appendix:degeneracy} for the relevant discussion. In particular, the amplitude of the $\tau$ derivative is notably smaller for P-ACT-L and P-ACT-LB compared to \textit{Planck} T\&E, reflecting the tighter constraint on $\tau$ provided by the compressed Gaussian likelihood of Ref.~\cite{Lee:2026toh}, which is constructed from the \texttt{Sroll2} likelihood. The derivatives for $\omega_b$ and $n_s$ are largely insensitive to the choice of dataset. Only the range $z \in [700, 1600]$ is shown for clarity; perturbations to $m_e(z)$ are considered over $z \in [300, 3000]$.}
\label{fig:deriv-Omega}
\end{figure*}

Figure \ref{fig:deriv-Omega} shows the functional derivatives $\delta\ln\Omega_{\rm BF}^i/\delta\ln m_e(z)$ for the three datasets, which characterize how the best-fit cosmological parameters respond to perturbations in $m_e(z)$. While the derivatives share the same qualitative oscillatory structure across all parameters and datasets, apparent differences in amplitude are visible for several parameters, most notably $\tau$, $\ln(10^{10}A_s)$, $\omega_c$, and $h$. These apparent differences, however, are not indicative of a physically distinct response of the recombination history to $m_e(z)$ across datasets, but rather reflect the different ways in which each dataset breaks standard CMB parameter degeneracies; we discuss this in detail in Appendix~\ref{appendix:degeneracy}.

\begin{figure}[ht!]
\centering
\includegraphics[width = \linewidth,trim= 00 20 00 0]{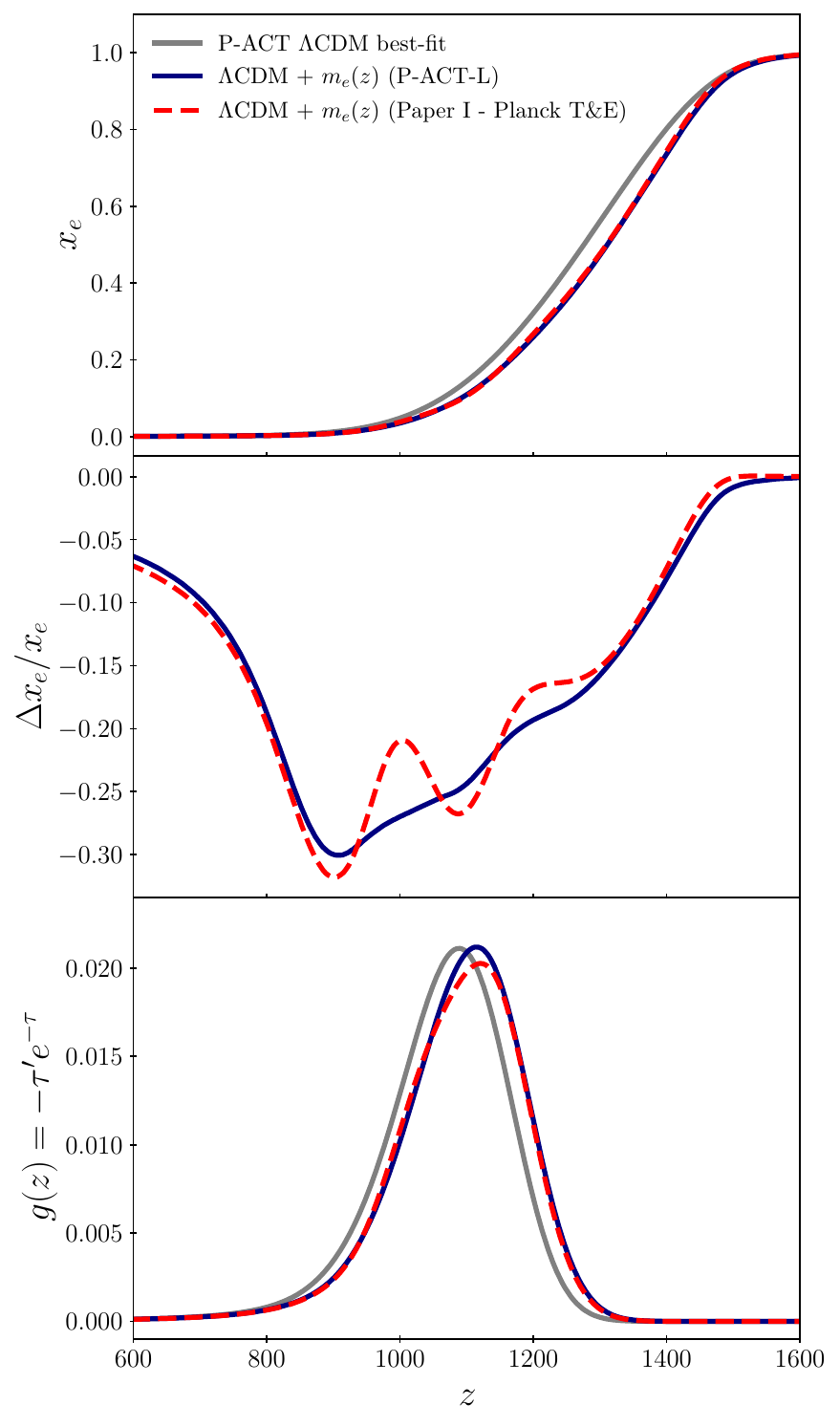}
\caption{Top: free-electron fraction $x_e(z)$ for the P-ACT $\Lambda$CDM best-fit \cite{AtacamaCosmologyTelescope:2025blo} (black) and the $\Lambda$CDM + $m_e(z)$ solutions targeting $H_0^{\rm target} = 73.04\;\mathrm{km\,s^{-1}\,Mpc^{-1}}$ with P-ACT-L (navy) and \textit{Planck} T\&E from Paper~\Romannum{1} (red dashed). Middle: fractional change $\Delta x_e/x_e$ relative to the $\Lambda$CDM best-fit. Bottom: the corresponding visibility functions $g(z) = -\tau'e^{-\tau}$ where the prime is the derivative with respect to the conformal time. Both $m_e(z)$ solutions produce qualitatively similar modifications to the recombination history, with the peak of the visibility function shifted to higher redshift compared to $\Lambda$CDM, consistent with the reduction of the sound horizon required to accommodate a higher $H_0$.}
\label{fig:P-ACT-L-xe-g}
\end{figure}

The corresponding changes in the free-electron fraction $x_e(z)$ and visibility function $g(z)$ are shown in Fig.~\ref{fig:P-ACT-L-xe-g}. The $m_e(z)$ solution constructed with P-ACT-L data produces a modified recombination history that is in close qualitative agreement with the Paper~\Romannum{1} solution obtained with \textit{Planck} T\&E data, with both exhibiting a suppression of $x_e$ at the level of $\sim\!20\%$ around recombination. The visibility function is slightly narrowed and its peak shifted to higher redshift compared to the standard $\Lambda$CDM prediction, consistent with an earlier recombination that reduces the sound horizon and yields a higher inferred value of $H_0$ from CMB data.

The best-fit cosmological parameters and $1\sigma$ uncertainties for the $\Lambda$CDM and $\Lambda$CDM + $m_e(z)$ models under P-ACT-L are given in Table~\ref{tab:sol}, both estimated using the Fisher matrix. With $m_e(z)$, the inferred $H_0 = 73.00 \pm 0.44\;\mathrm{km\,s^{-1}\,Mpc^{-1}}$ is fully consistent with the SH0ES measurement, resolving the tension. The shifts in the other cosmological parameters follow the same pattern as in Paper~\Romannum{1}: $\omega_c$ and $\omega_b$ both increase modestly, while $n_s$ decreases. The uncertainties of all parameters are in good agreement with the ACT DR6 constraints, confirming that our likelihood setup is faithful to the ACT data.

As a byproduct, and again in agreement with Paper~\Romannum{1}, the solution eases the $S_8$ tension: the inferred $S_8 = 0.800 \pm 0.010$\footnote{Since $S_8$ is a derived parameter, its uncertainty is estimated via linear error propagation through the Fisher matrix, as described in Appendix~\ref{appendix:derived}.} under $\Lambda$CDM + $m_e(z)$, compared to $S_8 = 0.830 \pm 0.010$ under $\Lambda$CDM, is in excellent agreement with the DES Year 6 cosmic shear measurement $S_8 = 0.798^{+0.014}_{-0.015}$ \cite{DES:2026mkc}.\footnote{We note, however, that the current status of the $S_8$ tension is subject to debate: KiDS Legacy finds consistency with the CMB at $<1\sigma$ ($S_8 = 0.815^{+0.016}_{-0.021}$ \cite{Wright:2025xka}), suggesting that survey-specific systematic effects may play a role in the discrepancy.} However, as in Paper~\Romannum{1}, the model is less consistent with late-time probes of $\Omega_m$: the increase in $H_0$ at fixed $\Omega_m h^3$, which controls the angular scale of the sound horizon, requires a decrease in $\Omega_m$, somewhat lower than the DESI DR2 constraint \cite{DESI:2025zgx} and inconsistent with the PantheonPlus constraint $\Omega_m = 0.334 \pm 0.018$ \cite{Brout:2022vxf}. This is the same fundamental limitation identified in Paper~\Romannum{1}, and motivates the inclusion of BAO data in the following subsection.

\begin{table}[ht!]
  \centering
  \begin{tabular}{c|c|c}
   Dataset &  \multicolumn{2}{c}{~~~P-ACT-L~~~}\\
    \hline
  Model & ~~~~~~~$\Lambda$CDM~~~~~~~
 & ~$\Lambda$CDM + $m_e(z)$~  \\
  \hline\hline
  $H_0$ (km/s/Mpc)   & $67.55\pm0.41$ & $73.00\pm0.44$ \\
    \hline
     $\omega_c$ & $0.1195\pm0.0010$ & $0.1234\pm0.0010$ \\
    \hline
     $100\omega_b$ & $2.249\pm0.011$ & $2.290\pm0.011$ \\
    \hline
     $\tau$ & $0.0588\pm0.0057$ & $0.0558\pm0.0061$ \\
    \hline
     $\ln(10^{10}A_s)$ & $3.051\pm0.010$ & $3.042\pm0.011$ \\
    \hline
     $n_s$ & $0.9698\pm0.0034$ & $0.9549\pm0.0035$ \\
     \hline\hline
     $S_8$ & $0.830\pm0.010$ & $0.800\pm0.010$ \\
     \hline
     $\Omega_m$ & $0.313\pm0.0058$ & $0.276\pm0.0051$
  \end{tabular}
  \caption{Best-fit values and $1\sigma$ uncertainties of the $\Lambda$CDM and $\Lambda$CDM + $m_e(z)$ models under our P-ACT-L dataset, estimated from the Fisher matrix. The $\Lambda$CDM + $m_e(z)$ model uses the time-varying electron mass constructed with $H_0^{\rm target} = 73.04\;{\rm km\,s^{-1}\,Mpc^{-1}}$ (black curve in Fig.~\ref{fig:P-ACT-L-solusions}). The $S_8$ uncertainty is estimated from error propagation within the Fisher matrix approximation (see Appendix~\ref{appendix:derived}).}
\vspace{-5mm}
\label{tab:sol}
\end{table}

\subsection{Application to P-ACT-LB data}

We now include DESI DR2 BAO data alongside the P-ACT-L CMB data to assess whether solutions consistent with all of these constraints can be found. The functional derivatives for this dataset are shown in Fig.~\ref{fig:deriv-Omega} alongside the P-ACT-L results. The inclusion of DESI DR2 visibly reduces the amplitude of the functional derivatives for $h$ and $\omega_c$, reflecting the tighter constraints these data place on the late-time expansion history and matter density.

We find that the maximum achievable $H_0$ with perturbative modifications to $m_e(z)$ is $H_0^{\rm BF} \approx 69.5\;\mathrm{km\,s^{-1}\,Mpc^{-1}}$, and the solutions approaching this upper bound become increasingly large in amplitude and oscillatory. This is qualitatively consistent with Paper~\Romannum{1}, and reflects the same fundamental limitation: raising $H_0$ through modifications to recombination requires lowering $\Omega_m$, which is tightly constrained by BAO independently of the CMB. The improved precision of DESI DR2 over BOSS DR12 leaves even less room for such accommodation, compared to what was found in Paper~\Romannum{1}.

\section{Discussion}
\label{sec:discussion}

In this work, we applied the Fisher-bias framework of Ref.~\cite{Lee:2022gzh} to assess what modifications to the cosmological recombination history are required to reconcile the $H_0$ inferred from CMB data with the value preferred by local distance ladder measurements, using the recent ACT DR6 CMB and DESI DR2 BAO datasets.

Our results with the P-ACT-L dataset closely parallel those found in Paper~\Romannum{1} using \textit{Planck} data alone. Notably, despite ACT DR6 providing more precise measurements of the CMB damping tail at small angular scales with independent instrumental systematics, a time-varying electron mass $m_e(z)$ can still be constructed that fully resolves the Hubble tension, raising the \textit{Planck}+ACT-inferred $H_0$ to the SH0ES value of $73.04\;\mathrm{km\,s^{-1}\,Mpc^{-1}}$ without worsening the fit to CMB data. This demonstrates that the recombination-based solution of Paper~\Romannum{1} is robust to the inclusion of a more precise and independent CMB dataset. While we have not included SPT-3G \cite{SPT-3G:2025bzu} data in our analysis, the addition of SPT-3G to \textit{Planck}+ACT provides only a modest improvement in CMB parameter constraints, and we do not expect it to qualitatively change our conclusions. We defer a full analysis including SPT-3G to future work.

The required $m_e(z)$ solution shares the same qualitative oscillatory structure found in Paper~\Romannum{1}. As a byproduct, the solution eases the $S_8$ tension as well, in agreement with Paper~\Romannum{1}. However, it increases $\omega_b$, which may be in tension with BBN constraints from primordial deuterium abundance measurements. Indeed, Ref.~\cite{Giovanetti:2026aku} recently argued that early-time solutions to the Hubble tension generically increase the preferred $\omega_b$, putting them in tension with BBN, and that including a BBN likelihood deters such models from recovering a high $H_0$. Consistently, the inferred $\omega_b = 0.02290 \pm 0.00011$ under $\Lambda$CDM + $m_e(z)$ is shifted upward relative to the BBN constraint $\omega_b = 0.02218 \pm 0.00055$ \cite{Schoneberg:2024ifp} at the $\sim\!1.3\sigma$ level. While a dedicated BBN analysis is beyond the scope of this work, this is an important caveat to keep in mind when assessing the viability of recombination-based solutions.

The situation changes substantially once late-time data are included. With the P-ACT-LB dataset, we find that perturbative modifications to $m_e(z)$ cannot achieve $H_0 \approx 70\;\mathrm{km\,s^{-1}\,Mpc^{-1}}$, and the solutions approaching this limit become increasingly non-perturbative and physically unappealing. This is qualitatively consistent with Paper~\Romannum{1}, and reflects the same fundamental limitation: raising $H_0$ through modifications to recombination generically lowers $\Omega_m$, which is further constrained by BAO independently of the CMB. Although DESI DR2 prefers a somewhat lower $\Omega_m$ than BOSS DR12, which naively favors recombination-based solutions, the improved precision of DESI DR2 more than compensates for this effect, making it even harder to raise $H_0$ through modifications to recombination.

This picture is consistent with a broader pattern seen across different approaches to the Hubble tension. Ref.~\cite{Lee:2025yah} found the same qualitative obstruction when modifying the primordial power spectrum: while such modifications can raise the CMB-inferred $H_0$ when only CMB data are considered, the required lowering of $\Omega_m$ renders those solutions incompatible with BAO and uncalibrated supernovae. The fundamental $H_0$--$\Omega_m$ incompatibility appears to be a generic feature of early-Universe solutions to the Hubble tension, once late-time data are included.

A complementary approach to assessing recombination-based solutions is the \texttt{ModRec} framework~\cite{Lynch:2024gmp}, which reconstructs the ionization history $x_e(z)$ directly from data using a model-agnostic parameterization and full MCMC, without invoking any physical mechanism that would drive changes in $x_e(z)$. While the two approaches ask fundamentally different questions --- \texttt{ModRec} reconstructs the posterior on $x_e(z)$ given the data, whereas our method finds the minimal modification to $m_e(z)$ achieving a specific target $H_0$ --- the conclusions are consistent. Applied to P-ACT-LB data, \texttt{ModRec} finds $H_0 = 69.6 \pm 1.0\;\mathrm{km\,s^{-1}\,Mpc^{-1}}$~\cite{AtacamaCosmologyTelescope:2025nti}, consistent with our finding that perturbative modifications to $m_e(z)$ cannot achieve $H_0 \approx 70\;\mathrm{km\,s^{-1}\,Mpc^{-1}}$ with the same dataset without deteriorating the fit, and finds no clear evidence for a deviation from the standard recombination history. We note that a time-varying $m_e(z)$ is not equivalent to a direct modification of $x_e(z)$ alone, since it also changes $\sigma_T \propto m_e^{-2}$, affecting CMB diffusion damping independently of $x_e(z)$. 

Our results therefore reinforce the conclusion of Paper~\Romannum{1}: perturbative modifications to the recombination history, while capable of fully resolving the Hubble tension with CMB data alone, cannot simultaneously satisfy the constraints from BAO, and uncalibrated supernovae. The addition of ACT DR6 and DESI DR2 data does not qualitatively change this conclusion, confirming its robustness across different CMB and BAO datasets. These results motivate extending this analysis to future CMB and BAO datasets, which will provide tighter constraints and further clarify the viability of recombination-based resolutions of the Hubble tension.

\section*{Acknowledgements}

We thank Marc Kamionkowski for useful discussions. N.\;L.\;was supported by the Horizon Fellowship from Johns Hopkins University. T.\;Z.\;was supported by a Haverford College KINSC Summer Scholars fellowship.

\begin{appendix}

\section{Likelihood-based numerical approach}
\label{appendix:method}

As noted in Sec.~\ref{sec:method}, directly forming the Gaussian $\chi^2$ requires explicit access to the covariance matrix of the data, which is not always straightforward to obtain for all likelihood functions. We therefore devise a numerical approach in which all required quantities are computed by taking numerical derivatives of the likelihood functions directly, without explicitly constructing the covariance matrix.

The key observation is that the quantities needed to evaluate Eqs.~\eqref{eq:DObf_X} and \eqref{eq:Dchi2bf_X} can be expressed in terms of the linear and quadratic responses of $\chi^2$ to shifts in the cosmological parameters and in $f(z)$. Specifically, Taylor-expanding $\chi^2$ around the fiducial cosmology to second order gives
\begin{eqnarray}
 \chi^2 &\approx& \chi^2(\mathbf{X}^{\rm fid})  + \mathbf{A}\cdot\left[\mathbf{X}(\vec{\Omega})-\mathbf{X}^{\rm fid}\right] \nonumber\\
 && + \frac{1}{2}\left[\mathbf{X}(\vec{\Omega})-\mathbf{X}^{\rm fid}\right] \cdot \mathbf{B} \cdot \left[\mathbf{X}(\vec{\Omega})-\mathbf{X}^{\rm fid}\right],
\end{eqnarray}
where $\mathbf{A}$ and $\mathbf{B}$ encode the linear and quadratic responses of the likelihood to changes in the theoretical prediction. The Fisher matrix and related quantities can then be written as
\begin{equation}
v_i \equiv \mathbf{A} \cdot \frac{\partial \mathbf{X}}{\partial \Omega^i}, 
\qquad
F_{ij} = \frac{1}{2}  \frac{\partial \mathbf{X}}{\partial \Omega^i} \cdot  \mathbf{B} \cdot \frac{\partial \mathbf{X}}{\partial \Omega^j},
\end{equation}
and similarly for the responses to $\Delta f(z)$:
\begin{eqnarray}
\tilde{v}_i &\equiv& \mathbf{A} \cdot \frac{\partial \mathbf{X}}{\partial f(z_i)}, 
\qquad
\widetilde{F}_{ij} \equiv \frac{1}{2} \frac{\partial \mathbf{X}}{\partial f(z_i)} \cdot \mathbf{B} \cdot \frac{\partial \mathbf{X}}{\partial f(z_j)},\nonumber\\
\lambda_{ij} &\equiv& \frac{\partial \mathbf{X}}{\partial \Omega^i} \cdot \mathbf{B} \cdot \frac{\partial \mathbf{X}}{\partial f(z_j)}.
\end{eqnarray}
The key point is that $\mathbf{A}$ and $\mathbf{B}$ need not be constructed explicitly. Instead, all required quantities can be obtained from finite differences of $\chi^2$ evaluated at the fiducial cosmology. Specifically, the linear response $v_i$ and the Fisher matrix elements $F_{ij}$ are obtained from numerical derivatives of $\chi^2$ with respect to the cosmological parameters $\Omega^i$:
\begin{align}
v_i &= \frac{\partial \chi^2}{\partial \Omega^i}\bigg|_{\rm fid}, \\
F_{ij} &= \frac{1}{2}\frac{\partial^2 \chi^2}{\partial \Omega^i \partial \Omega^j}\bigg|_{\rm fid}.
\end{align}
Similarly, $\tilde{v}_i$, $\widetilde{F}_{ij}$, and $\lambda_{ij}$ are obtained from numerical derivatives of $\chi^2$ with respect to $f(z_i)$ and mixed derivatives with respect to $\Omega^i$ and $f(z_j)$. This approach is applicable to any likelihood that is well approximated as Gaussian around the fiducial cosmology.

Once these quantities are computed, the shifts in the best-fit parameters and chi-squared due to $\Delta f(z)$ follow from minimizing $\chi^2$ with respect to $\Delta\Omega^i$, giving
\begin{equation}
\Delta \Omega^i_{\rm BF} = -\frac{1}{2}(F^{-1})_{ij}\lambda_{jk}\Delta f_k,
\label{eq:Omegabf-app}
\end{equation}
and
\begin{eqnarray}
\Delta \chi^2_{\rm BF} &=& \left(\tilde{v}_i - \frac{1}{2}v_j(F^{-1})_{jk}\lambda_{ki}\right)\Delta f_i \nonumber\\
&&+ \Delta f_i \left(\widetilde{F}_{ij} - \frac{1}{4}\lambda_{qi}(F^{-1})_{qk}\lambda_{kj}\right) \Delta f_j,
\label{eq:Dchi2bf-app}
\end{eqnarray}
which are the main results used in the numerical analysis. For any arbitrary function $\Delta f(z)$, these expressions provide the expected shift in the best-fit cosmological parameters and the corresponding change in best-fit $\chi^2$, allowing us to solve the optimization problem of Eq.~\eqref{eq:minimization}. One can verify that Eqs.~\eqref{eq:Omegabf-app} and \eqref{eq:Dchi2bf-app} are equivalent to the functional derivative expressions given in Sec.~\ref{sec:method}.

\section{Functional derivative of\\the best-fit chi-squared}
\label{appendix:dchi2}

\begin{figure*}[ht!]
\centering
\includegraphics[width = .45\linewidth,trim= 0 0 0 0]{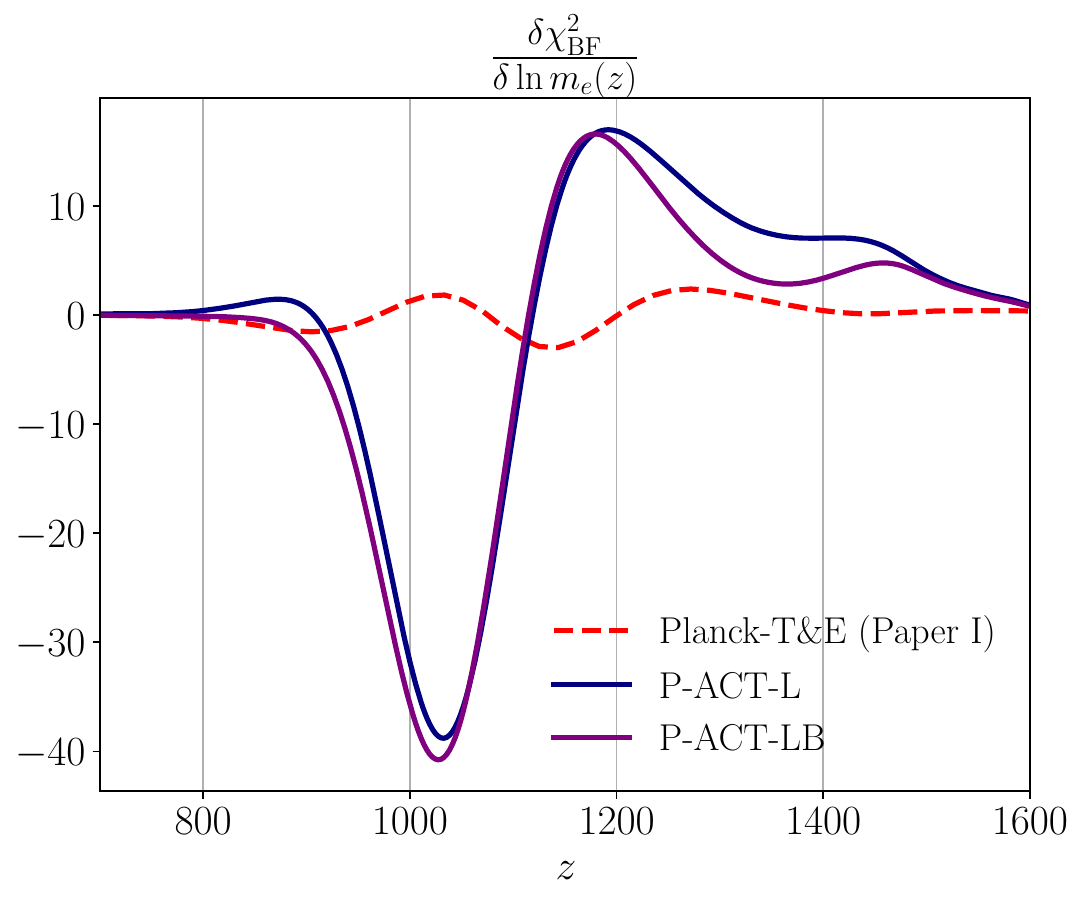}\\
\includegraphics[width = .44\linewidth,trim= 0 20 0 0]{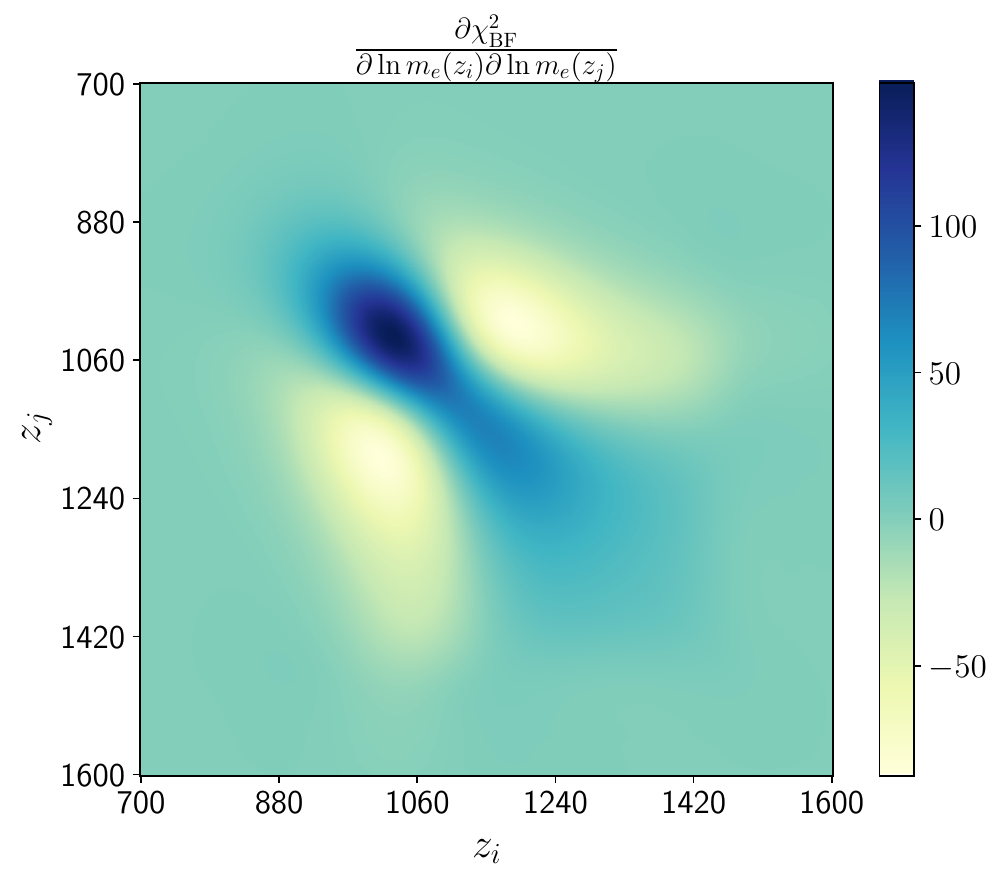}
\includegraphics[width = .43\linewidth,trim= 0 20 0 0]{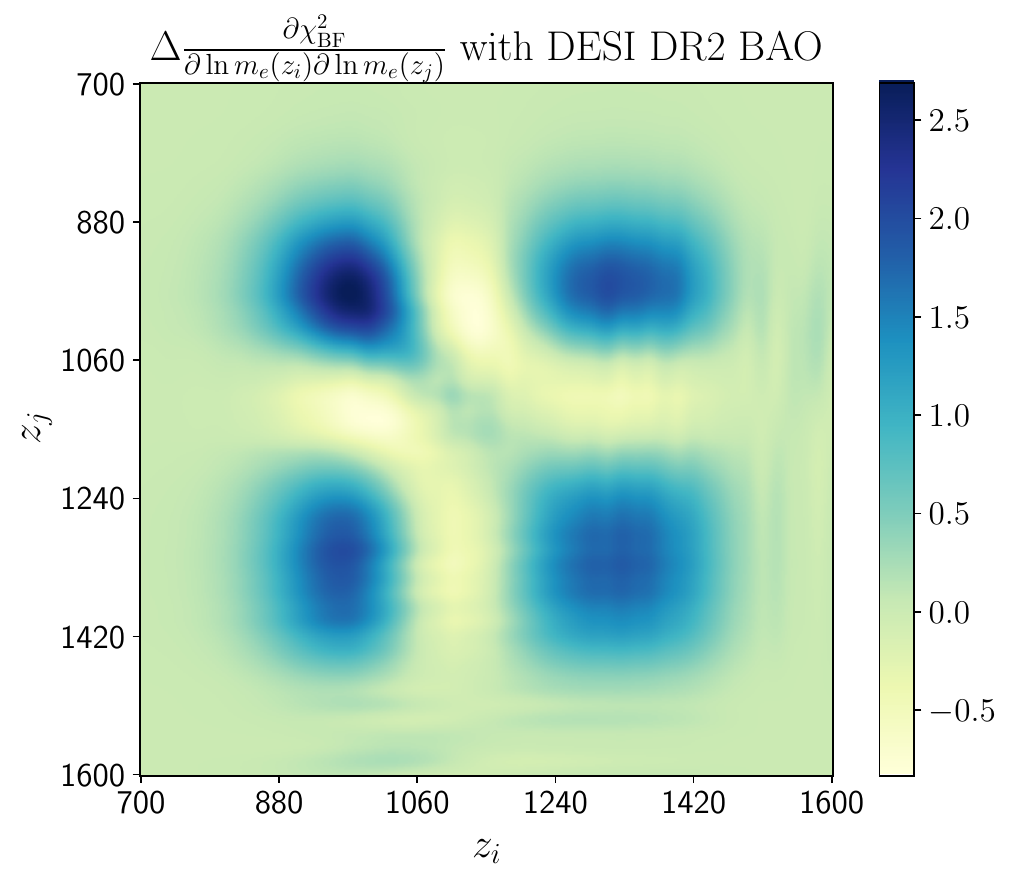}
\caption{Top: linear functional derivative of the best-fit chi-squared $\delta\chi^2_{\rm BF}/\delta\ln m_e(z)$ for \textit{Planck} T\&E (red dashed, Paper~\Romannum{1}), P-ACT-L (navy), and P-ACT-LB (purple). The P-ACT-L and P-ACT-LB derivatives show notably larger amplitudes than \textit{Planck} T\&E. Bottom left: quadratic response of the best-fit chi-squared $\partial^2\chi^2_{\rm BF}/\partial\ln m_e(z_i)\partial\ln m_e(z_j)$ with respect to logarithmic changes in electron mass, given P-ACT-L data. Bottom right: additional contribution to the quadratic response when DESI DR2 are included. The quadratic response is also notably larger in amplitude than in Paper~\Romannum{1}, consistent with the large amplitude of the linear derivative. Despite these large individual amplitudes, the resulting $m_e(z)$ solutions are very similar across datasets, suggesting large cancellations between the linear and quadratic contributions to $\Delta\chi^2_{\rm BF}$.}
\label{fig:dchi2}
\end{figure*}

Figure \ref{fig:dchi2} shows the linear functional derivative $\delta\chi^2_{\rm BF}/\delta\ln m_e(z)$ and the quadratic response $\partial^2\chi^2_{\rm BF}/\partial\ln m_e(z_i)\partial\ln m_e(z_j)$ for the three datasets. Both the linear and quadratic derivatives show notably larger amplitudes for P-ACT-L compared to \textit{Planck} T\&E, by roughly an order of magnitude (see Fig.~5 of Paper~\Romannum{1} for the quadratic derivatives with \textit{Planck} T\&E dataset). The additional contribution from DESI DR2 to the quadratic response (bottom right) is much smaller in amplitude, consistent with the fact that BAO data constrain background quantities at late times and have limited sensitivity to the recombination history.

\onecolumngrid

\vspace{0.1in}

\twocolumngrid

A direct inference of the solution shape from the linear functional derivative $\delta\chi^2_{\rm BF}/\delta\ln m_e(z)$ alone is not straightforward, since the solution is determined by minimizing $\|\Delta f\|^2$ subject to both the linear and quadratic responses of $\chi^2_{\rm BF}$. Indeed, despite the notably larger amplitudes of both derivatives for P-ACT-L, the resulting $m_e(z)$ solutions are very similar across datasets. This is likely due to large cancellations between the linear and quadratic contributions to $\Delta\chi^2_{\rm BF}$ in the P-ACT-L case, such that their sum, which is what the constraint $\Delta\chi^2_{\rm BF} \leq 0$ directly involves, remains comparable to the \textit{Planck} T\&E case.

\section{Parameter degeneracies and\\functional derivatives}
\label{appendix:degeneracy}

\begin{figure}[ht!]
\centering
\includegraphics[width = \linewidth,trim= 00 20 00 0]{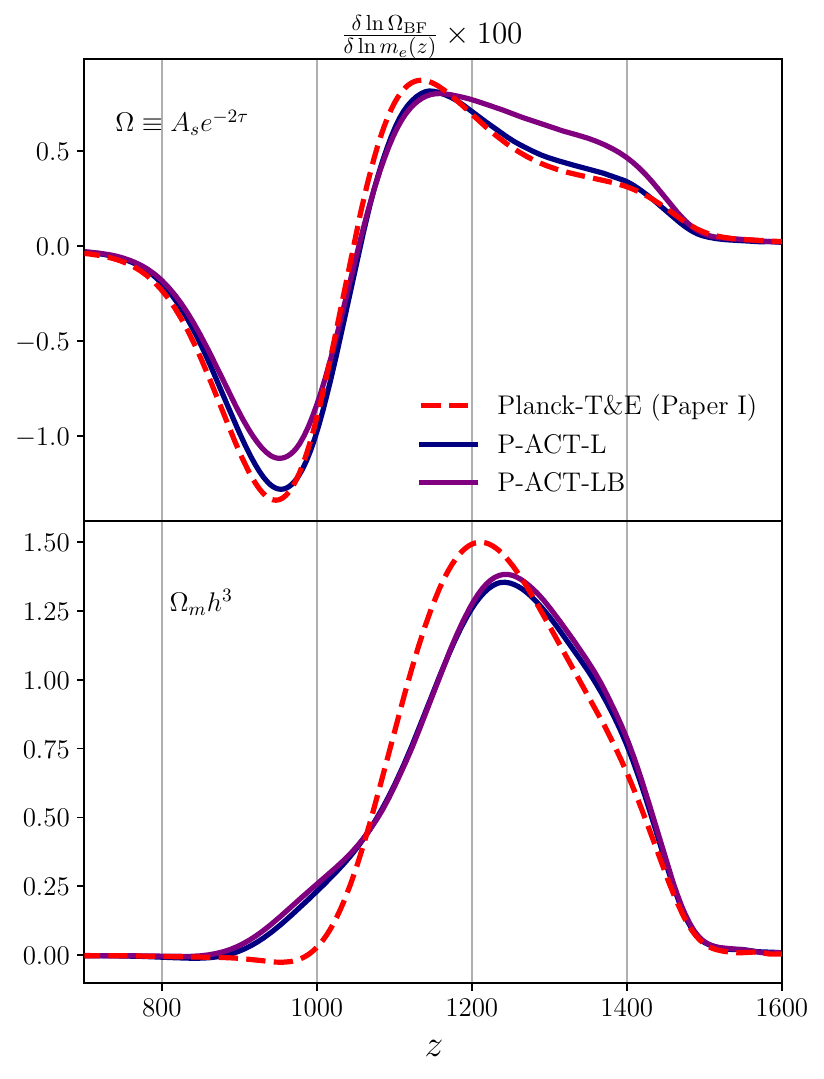}
\caption{Functional derivatives of the degenerate parameter combinations $A_s e^{-2\tau}$ (top) and $\Omega_m h^3$ (bottom) with respect to $\ln m_e(z)$, for \textit{Planck} T\&E (red dashed, Paper~\Romannum{1}), P-ACT-L (navy), and P-ACT-LB (purple). Top: the three curves are in good agreement, demonstrating that the apparent differences in the individual $\tau$ and $\ln(10^{10}A_s)$ derivatives seen in Fig.~\ref{fig:deriv-Omega} are a consequence of the $\tau$--$A_s$ degeneracy being resolved differently by each dataset. 
The small residual difference between P-ACT-L and P-ACT-LB is an indirect effect propagated through parameter degeneracies in the Fisher matrix. Bottom: in contrast, the \textit{Planck} T\&E derivative shows a notably larger amplitude and does not extend to $z \lesssim 1000$, while P-ACT-L and P-ACT-LB agree closely with each other. This reflects genuine additional constraining power of ACT DR6 on $\Omega_m h^3$ through its sensitivity to the low-redshift damping tail, whereas DESI DR2, constraining background quantities, do not directly affect this combination.}
\label{fig:degeneracy}
\end{figure}

The functional derivatives $\delta\ln\Omega_{\rm BF}^i/\delta\ln m_e(z)$ shown in Fig.~\ref{fig:deriv-Omega} exhibit apparent differences in amplitude across datasets for several parameters, most notably $\tau$, $\ln(10^{10}A_s)$, $\omega_c$, and $h$. Here we show that these differences are not indicative of any physically distinct sensitivity of the recombination history to $m_e(z)$ across datasets, but rather reflect the different ways in which each dataset breaks standard CMB parameter degeneracies. In Fig.~\ref{fig:degeneracy}, we show the functional derivatives of two degenerate combinations, $A_s e^{-2\tau}$ and $\Omega_m h^3$, which we discuss in turn.\\

\textbf{The $\tau$--$A_s$ degeneracy.}
The parameters $\tau$ and $A_s$ are degenerate through the combination $A_s e^{-2\tau}$, which controls the overall amplitude of the CMB power spectra. When each dataset breaks this degeneracy differently, the individual $\tau$ and $A_s$ derivatives can differ substantially even if the underlying response of the recombination history is the same. The top panel of Fig.~\ref{fig:degeneracy} shows the functional derivative of this combination for all three datasets. The three curves are in good agreement, confirming that the apparent differences in the individual $\tau$ and $\ln(10^{10}A_s)$ derivatives shown in Fig.~\ref{fig:deriv-Omega} are a consequence of the $\tau$--$A_s$ degeneracy being resolved differently by each dataset rather than any physically distinct sensitivity to $m_e(z)$. The small residual difference between P-ACT-L and P-ACT-LB is understood as an indirect effect propagated through parameter degeneracies in the Fisher matrix: while DESI DR2 constrains background quantities and has no direct sensitivity to perturbation amplitudes, tightening the constraints on $\Omega_m$ and $h$ shifts the degeneracy directions among all six cosmological parameters through $(F^{-1})_{ij}$, inducing small indirect changes in combinations such as $A_s e^{-2\tau}$.\\

\textbf{The $\Omega_m h^3$ combination.}
The combination $\Omega_m h^3$ controls the angular scale of the sound horizon $\theta_s$ and is therefore tightly constrained by the CMB independently of late-time data~\cite{Lee:2022gzh}. The bottom panel of Fig.~\ref{fig:degeneracy} shows the functional derivative of this combination for the three datasets. In contrast to the $A_s e^{-2\tau}$ case, a noticeable difference between the \textit{Planck} T\&E and P-ACT-L derivatives is visible. Most notably, the P-ACT-L derivative extends to lower redshifts compared to \textit{Planck} T\&E, with a visible contribution already at $z \lesssim 1000$, whereas the \textit{Planck} T\&E derivative remains close to zero there. This reflects the fact that ACT data are more sensitive to the low-redshift damping tail, probing angular scales where recombination physics at $z \lesssim 1000$ leaves characteristic imprints, thereby providing genuine additional constraining power on $\Omega_m h^3$ 
beyond what \textit{Planck} captures. In contrast, the P-ACT-L and P-ACT-LB derivatives agree closely with each other, confirming that DESI DR2 does not directly affect the $\Omega_m h^3$ combination, as expected since BAO data constrain background quantities with no direct sensitivity to $\Omega_m h^3$ through recombination physics.

\section{Uncertainty estimation for\\derived parameters}
\label{appendix:derived}

In this work, we quote best-fit values and $1\sigma$ uncertainties for derived parameters such as $S_8 \equiv \sigma_8(\Omega_m/0.3)^{0.5}$, which are not directly varied in the Fisher-bias formalism but are functions of the six base $\Lambda$CDM parameters $\vec{\Omega} = \{\omega_b, \omega_c, h, \tau, \ln(10^{10}A_s), n_s\}$. Here we describe how these are estimated consistently within the Fisher matrix approximation.\\

\textbf{Best-fit value.}
The shift in any derived parameter $q = q(\vec{\Omega})$ away from its fiducial value follows directly from the shifts in the base parameters,
\be
\Delta q^{\rm BF} = \sum_i \frac{\partial q}{\partial \Omega^i} \bigg|_{\vec{\Omega}_{\rm fid}} \Delta \Omega^i_{\rm BF},
\ee
where $\Delta \Omega^i_{\rm BF}$ are the best-fit shifts obtained from the Fisher-bias formalism, Eq.~\eqref{eq:DObf_X}. This is a Taylor expansion around the fiducial cosmology, valid within the same perturbative regime assumed throughout the main analysis.

{\boldmath $1\sigma$} \textbf{uncertainty.}
Within the Fisher matrix approximation, the posterior on $\vec{\Omega}$ is Gaussian with covariance $F^{-1}$,
\be
\mathcal{L}(\vec{\Omega}) \propto \exp\left(-\frac{1}{2} (\vec{\Omega}-\vec{\Omega}_{\rm BF})^T F\; (\vec{\Omega}-\vec{\Omega}_{\rm BF})\right).
\ee
Linearizing $q$ around $\vec{\Omega}_{\rm BF}$ as above, $q$ inherits a Gaussian distribution from the Gaussian posterior on $\vec{\Omega}$. Its variance is given by standard linear error propagation,
\be
\sigma_q^2 = \sum_{i,j} \frac{\partial q}{\partial \Omega^i}(F^{-1})_{ij}\frac{\partial q}{\partial \Omega^j},
\ee
which follows from $\text{Cov}(\delta\Omega^i, \delta\Omega^j) = (F^{-1})_{ij}$ for a Gaussian with precision matrix $F$. The only additional assumption beyond the Fisher matrix approximation is that $q$ is well approximated as a linear function of $\vec{\Omega}$ over the width of the posterior. Given that the Fisher matrix uncertainties on the base parameters are at the percent level or below, this linearization is an excellent approximation for smooth derived parameters such as $S_8$.

The partial derivatives $\partial q / \partial \Omega^i$ are computed numerically using two-sided finite differences around the fiducial cosmology with \texttt{CLASS}. All ingredients --- $\Delta\Omega^i_{\rm BF}$, $F^{-1}$, and $\partial q/\partial\Omega^i$ --- are already available from the main analysis, so this estimate adds no significant computational cost.

\end{appendix}

\bibliography{mybib}

\end{document}